\documentclass[3p,times]{elsarticle}

\usepackage{ecrc}
\usepackage{subfigure}

\volume{00}

\firstpage{1}

\journalname{Nuclear Physics A}

\runauth{}

\jid{npa}

\jnltitlelogo{Nuclear Physics A}

\usepackage{amssymb}
\biboptions{square,comma,numbers,sort&compress}

\usepackage[figuresright]{rotating}
\usepackage{xcolor}

\begin{document}

\begin{frontmatter}



% \dochead{}
%% e.g. \dochead{17th International Conference on Dynamical Processes in Excited States of Solids}
\ead{marlene.nahrgang@phy.duke.edu}
\title{D mesons in non-central heavy-ion collisions:\\ fluctuating vs. averaged initial conditions}

\author{Marlene Nahrgang$^{1,2}$, J\"org Aichelin$^{3}$, Pol Bernard Gossiaux$^{3}$, Klaus Werner$^{3}$}

\address{$^1$ Department of Physics, Duke University, Durham, North Carolina 27708-0305, USA}
\address{$^2$ Frankfurt Institute for Advanced Studies (FIAS), Ruth-Moufang-Str.~1, 60438 Frankfurt am Main, Germany}
\address{$^3$ SUBATECH, UMR 6457, Universit\'e de Nantes, Ecole des Mines de Nantes,
IN2P3/CNRS. 4 rue Alfred Kastler, 44307 Nantes cedex 3, France}

% \address{}

\begin{abstract}
The suppression of D mesons in non-central heavy-ion collisions is investigated. The anisotropy in collisions at finite impact parameter leads to an ordering of all-angle, in- and out-of-plane nuclear modification factors due to the different in-medium path lengths. Within our MC@sHQ+EPOS model of heavy-quark propagation in the QGP we demonstrate that fluctuating initial conditions lead to an effective reduction of the energy loss of heavy quarks, which is seen in a larger nuclear modification factor at intermediate and high transverse momenta. The elliptic flow at small transverse momenta is reduced. 
\end{abstract}


\end{frontmatter}

% \linenumbers

\section{Introduction}
Heavy-quark observables like the nuclear modification factor $R_{\rm AA}$ and the elliptic flow $v_2$ are measured in ultra-relativistic heavy-ion collisions at RHIC \cite{rhic} and the LHC \cite{Alice,delValle:2012qw,Abelev:2013lca}. The results indicate a substantial medium modification of the spectra: a suppression at high transverse momentum $p_T$ and partial thermalization with the QGP medium at low $p_T$. Many theoretical approaches describing the energy loss via elastic scatterings  \cite{elastic}, gluon bremsstrahlung 
\cite{radiative}
  or a mixture of both processes exist. For some of these approaches, it is necessary to rescale these interaction cross sections or the diffusion coefficient and to couple to a background fluid dynamical medium of light partons in order to reproduce the experimental data within numerical simulations  \cite{results,Gossiaux:2008jv,Gossiaux:2010yx,Gossiaux:2012ya}. It remains, however, a challenge to describe $R_{\rm AA}$ and $v_2$ within a single setup.

In order to discriminate between the various models it becomes therefore necessary to improve the description of the fluid dynamical evolution, explore the potential of new heavy-quark observables, such as azimuthal correlations \cite{Nahrgang:2013saa, Nahrgang:2013pka}, and to study the centrality and mass dependence of these observables. In this work, we present the coupled MC@sHQ+EPOS model and study the $R_{\rm AA}$ in non-central heavy-ion collisions as all-angle, in- and out-of-plane observables. In particular, we discuss the influence of fluctuating versus smooth initial conditions. This latter aspect was studied in the frame-work of (light-hadron) jet physics \cite{Rodriguez:2010di,Renk:2011qi,Zhang:2012ik}, but only recently attracted attention in the  context of heavy-quark observables \cite{Cao:2014fna}, where contrary to the present work it is found that the energy loss of heavy quarks is enhanced in a medium with local fluctuations.

% \label{}

% \newpage

 \section{MC@sHQ + EPOS}
Recently, we presented first results from our Monte-Carlo heavy quark propagation, MC@sHQ, coupled to the state-of-the-art fluid dynamical evolution of the QGP from EPOS initial conditions \cite{Nahrgang:2013xaa, Nahrgang:2013saa, Nahrgang:2013pka}. Three main stages of the coupled evolution are described: 1) initialization at the nucleon-nucleon collision points, thus inside the initial hot spots, according to the FONLL momentum distribution \cite{fonll},  2) propagation through the medium by Monte-Carlo sampling of the Boltzmann equation, and 3) hadronization at $T=155$~MeV according to coalescence and fragmentation.

The cross sections in the Boltzmann equation are either given by purely elastic scatterings off the gluons and light quarks, drawn from the thermal medium, or include radiative corrections. Both interaction scenarios are investigated in the present work. The respective scattering rates are rescaled by a $K$-factor in order to reproduce the high-$p_T$ $R_{\rm AA}$ data for D mesons measured by ALICE \cite{delValle:2012qw} in central collisions. All other heavy-quark observables, like the elliptic flow and azimuthal correlations and their centrality or mass dependence, are calculated with the same $K$-factor. With $K=0.8$ for the collisional+radiative interaction mechanism we are rather close to the generic case of $K=1.0$.

It  had been shown that different description of the background medium can lead to large uncertainties in the $R_{\rm AA}$ of D mesons \cite{Gossiaux:2011ea}. It is, therefore, important to use a model for the light hadron sector, which compares successfully to experimental data. The EPOS model is capable of describing transverse momentum spectra, flow harmonics and correlation patterns like the ridge in Au+Au collisions at RHIC and Pb+Pb and p+Pb collisions at the LHC \cite{EPOS}.

In principle, the EPOS fluctuating initial conditions allow us to study heavy quark observables in an event-by-event setup. For the numerical simulation, however, we run $N^{\rm HQ}$ events per EPOS event in order to be more efficient in computing power. The particular choice of $N^{\rm HQ}$ will be discussed in this work.

We note here, that the current version of MC@sHQ+EPOS does not include shadowing, which is expected to suppress the initial heavy quark momentum spectrum at low $p_T$ as compared to the proton-proton reference and thus reduces the low-$p_T$ $R_{\rm AA}$ even in the absence of the QGP. Shadowing will be included in the next version.

 \section{Results}

\begin{figure}
 \centering
  \subfigure{\label{fig:raacoll1}\includegraphics[width=0.49\textwidth]{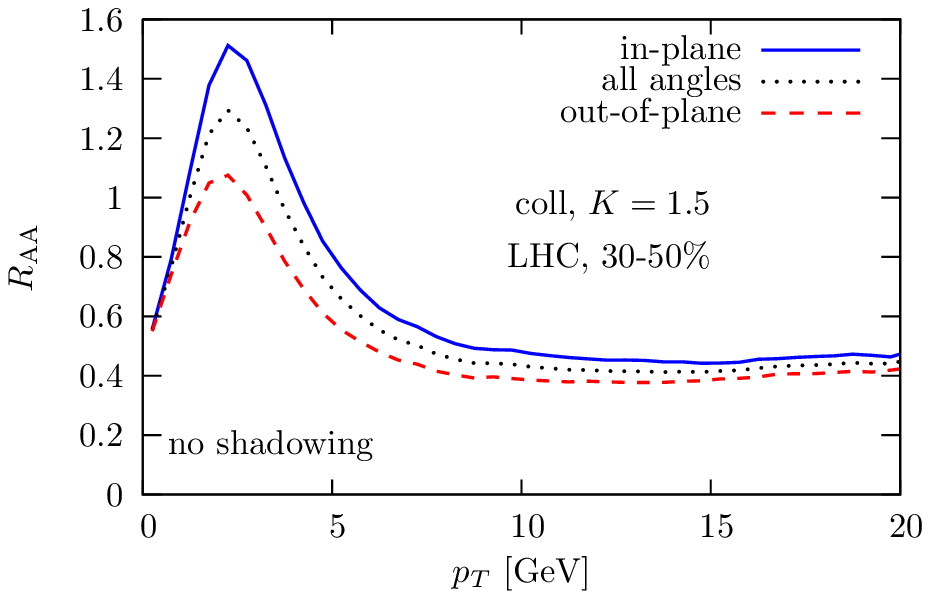}}\hfill
  \subfigure{\label{fig:raaradcoll1}\includegraphics[width=0.49\textwidth]{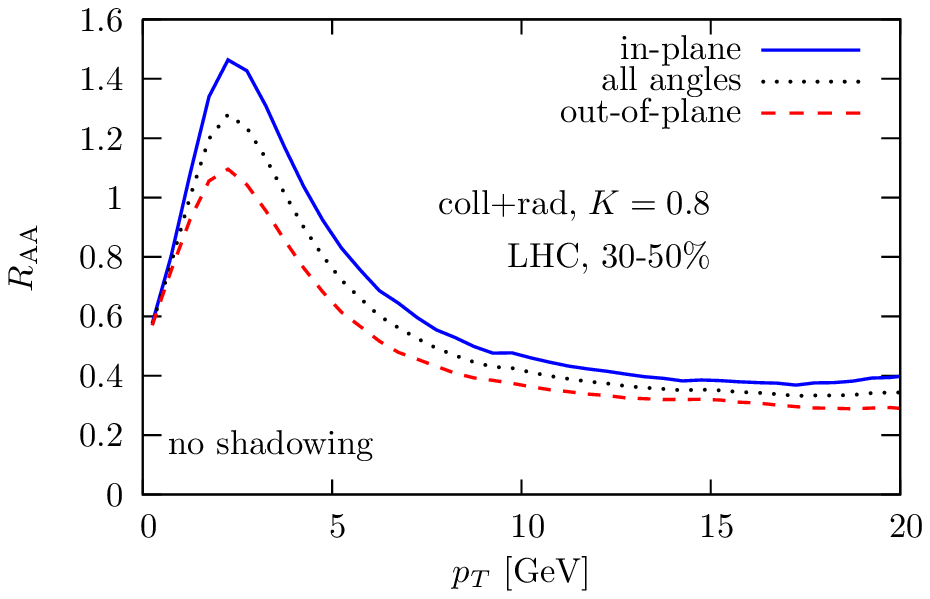}}
 \caption{The in- and out-of-plane $R_{\rm AA}$ of D mesons in the $30-50$~\% centrality class in comparison to the all-angle $R_{\rm AA}$ for the purely collisional (left) and the collisional+radiative (right) interaction mechanism for fluctuating initial conditions.}
 \label{fig:raa1}
\end{figure}

The $R_{\rm AA}$ of D mesons in $30-50$\% most central heavy-ion collisions at $\sqrt{s}=2.76$~TeV is determined in the full azimuth over all angles and separately in in-plane and out-of-plane directions. The in-plane $R_{\rm AA}$ is defined for D mesons with $\Delta\phi\in[-\pi/4,\pi/4]$ and $[3\pi/4,5\pi/4]$, where $\Delta\phi$ is the azimuthal angle with respect to the participant plane. The out-of-plane $R_{\rm AA}$ is defined accordingly with $\Delta\phi\in[\pi/4,3\pi/4]$ and $[5\pi/4,7\pi/4]$. Due to the initial spatial anisotropy high-$p_T$ heavy quarks which initially go in-plane traverse less QGP matter than those initially going out-of-plane. In Fig. \ref{fig:raa1} we observe the expected ordering of the suppression for $1$~GeV$<p_T<20$~GeV. Both interaction scenarios give very similar $R_{\rm AA}$ over the shown $p_T$ range. The rising trend at high $p_T$ in the purely collisional case can be seen, while the $R_{\rm AA}$ for the scenario with radiative corrections seems to saturate toward high $p_T$.

Due to the lack of shadowing and the strong radial flow in the EPOS environment our calculations for the $R_{\rm AA}$ are above the data for the lower $p_T$ in the $0-20$\% most central collsions. We expect this issue to persist also in the non-central collisions investigated here.

\begin{figure}
 \centering
  \subfigure{\label{fig:inicond_av}\includegraphics[width=0.35\textwidth]{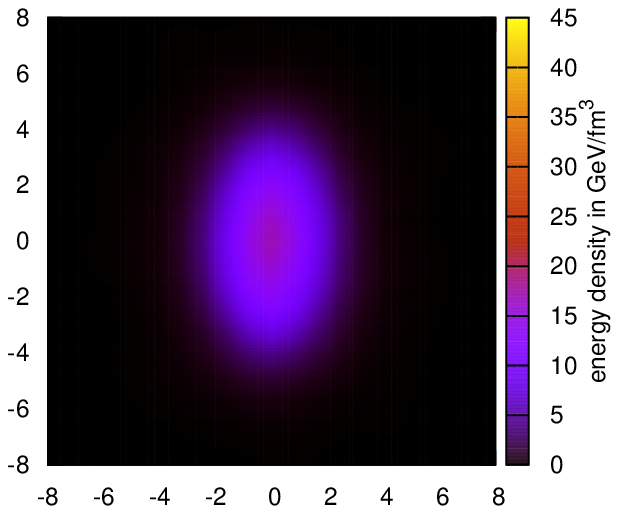}}\hfil
  \subfigure{\label{fig:inicond_fluc}\includegraphics[width=0.35\textwidth]{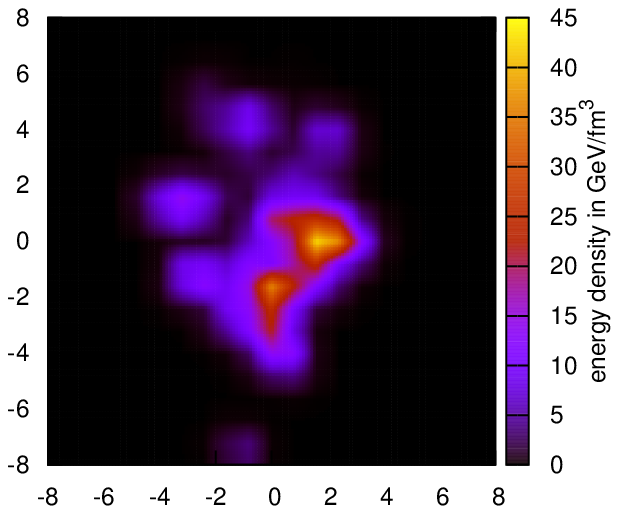}}
 \caption{Initial energy density for $30-50$\% most central events in Pb+Pb collisions at $\sqrt{s}=2.76$~TeV. The left plot shows the average over the initial conditions of $100$ events. The right plot is the energy density of one random event.}
 \label{fig:inicond}
\end{figure}

We have the possibility to study the heavy-quark propagation in a fluid dynamical environment, which is subsequent to either (event-by-event) fluctuating or (event-) averaged initial conditions. In fig. \ref{fig:inicond} we show the energy density profiles at midrapidity ($\eta=0$) for initial conditions, which are averaged over $1000$ EPOS initial stages of Pb+Pb collisions in the 30-50\% centrality class (left), and for one random event in the same centrality class (right). In both cases the x-axis is aligned with the angle of the participant plane. The elliptic initial anisotropy is well observed for the averaged initial conditions, which produce a smooth profile. In the initial conditions for a single event, however, one sees denser/hotter spots as well as more dilute/colder spots. The peak energy density in the hot spots is much higher than the averaged values. 

The following two aspects are crucial for investigating the energy loss in a fluctuating medium. 1) The correlation of the production points of the heavy quarks and the initial hot spots, which leads locally to an enhanced energy loss as compared to a scenario with averaged initial conditions. 2) The increase of the heavy-quark energy loss with the energy density of the medium, which is less than linear. In order to elaborate on this last point, one can look at a medium with local fluctuations $\varepsilon(x)=\varepsilon_0+\delta\varepsilon(x)$ around an averaged energy density $\langle\varepsilon(x)\rangle=\varepsilon_0$. The local energy loss is related to the temperature by ${\rm d}E/{\rm d}x\propto T^\beta$
with $\beta$ ranging from $\approx 1$ for collisional energy loss with running $\alpha_s$ to $\beta\approx 2$ for radiative LPM \cite{Gossiaux:2012cv}, and $\varepsilon \propto T^n$, with $n\approx 4$. Thus 
\begin{equation}
\frac{{\rm d}E}{{\rm d}x}\propto \varepsilon^\delta\quad{\rm with}\quad \delta=\frac{\beta}{n}<1\,.
\label{eq:dedx}
\end{equation} 
A Taylor expansion around $\varepsilon_0$ gives an averaged energy loss of 
\begin{equation}
\left\langle \frac{{\rm d}E}{{\rm d}x} \right\rangle = \frac{{\rm d}E(\varepsilon_0)}{{\rm d}x}
\times \left[1 - \frac{\delta(1-\delta)}{2} \frac{\langle \delta \varepsilon^2\rangle}{\varepsilon_0^2}+\cdots\right]\,.
\end{equation} 
We see that due to the negative curvature of ${\rm d}E/{\rm d}x$ as a function of the energy density ($\delta<1$), the effective energy loss of the heavy quarks in a fluctuating medium is reduced compared to the energy loss in an averaged medium. 

\begin{figure}
 \centering
  \subfigure{\label{fig:RAAinicond}\includegraphics[width=0.5\textwidth]{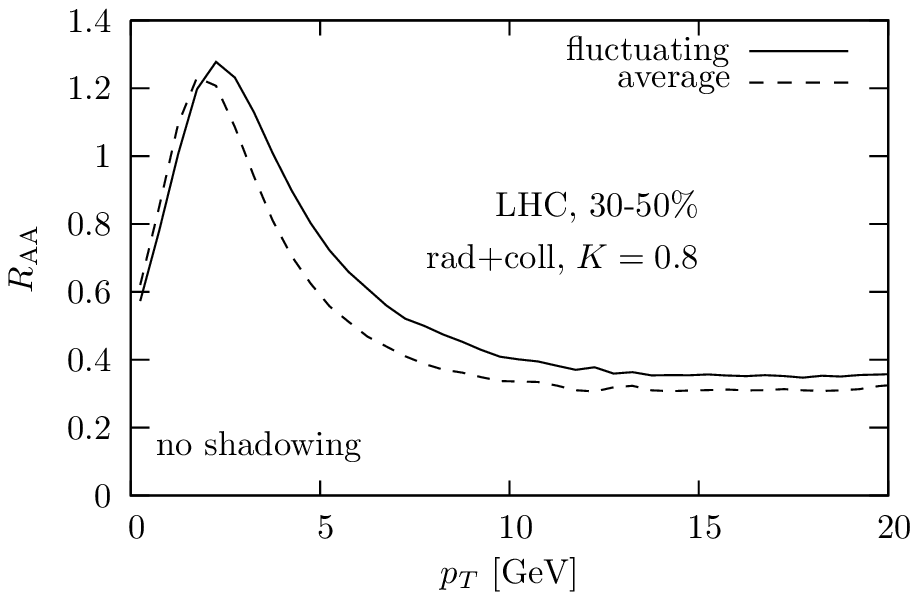}}\hfill
  \subfigure{\label{fig:v2inicond}\includegraphics[width=0.5\textwidth]{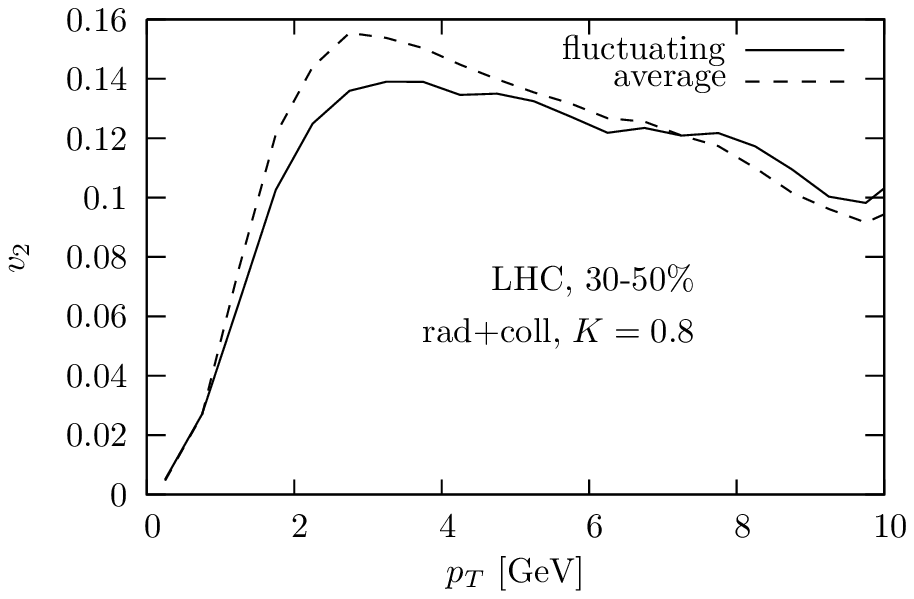}}
 \caption{The $R_{\rm AA}$ (left) and the $v_2$ (right) compared for fluctuating and averaged initial conditions.}
 \label{fig:inicondobs}
\end{figure}

In our numerical simulations for an expanding medium we see that this second effect, where the fluctuations in the medium reduce the energy loss, dominates. In Fig.~\ref{fig:inicondobs} we present results for the  $R_{\rm AA}$ (left) and the $v_2$ (right) for the scenario of collisional plus radiative energy loss. 
The factor $K=0.8$ is tuned to reproduce the $R_{\rm AA}$ for higher $p_T$ in central collisions in the case of fluctuating initial conditions. Since we are interested in the mere influence of the initial conditions we do not recalibrate the $K$-factor for energy loss in the case of averaged initial conditions. In the $R_{\rm AA}$ (left) one observes a smaller quenching in the fluctuating case for all transverse momenta $p_T>3$~GeV. In central collisions we expect to find the same ordering: larger quenching for averaged than for fluctuating initial conditions.

The result for the $R_{\rm AA}$ is largely independent of the number $N^{\rm HQ}$ of MC@sHQ events, which we run per EPOS fluid dynamical event, and results converge quickly with the number $N^{\rm EPOS}$ of EPOS fluid dynamical events, when the overall statistics $N^{\rm HQ}\times N^{\rm EPOS}$ is high enough. In the present study, $N^{\rm HQ}=10^4$.

In fig.~\ref{fig:inicondobs} (right) we show the $v_2$ results for the same evolution scenarios obtained from correlations of the heavy quarks with the participant plane. At low momenta $p_T\lesssim 5$~GeV we find that the averaged initial conditions lead to a larger $v_2$ than the fluctuating initial conditions. The hot and dense spots in the initial energy density are themselves rather spherical, which reduces the spatial anisotropy. Local pressure gradients in these regions produce an azimuthally isotropic expansion. This reduces the overall elliptic flow. At larger $p_T$ the elliptic flow is built significantly from energy loss along the different path lengths in- and out-of-plane. Here a more detailed study of these contributions with enhanced statistics is mandatory. 

For the elliptic flow results to converge one needs to run a sufficiently large number of EPOS fluid dynamical events $N^{\rm EPOS}$. Due to computational limitations, this can be achieved by running $N^{\rm HQ}<=10^4$ heavy quark events per one EPOS fluid dynamical event. For the study of the elliptic flow only initial fluctuations are taken into account. The event-by-event fluctuations of the final event plane in comparison to the initial participant plane are not considered here. It is not expected that shadowing affects the $v_2$.

\begin{figure}
 \centering
  \subfigure{\label{fig:correl1}\includegraphics[width=0.48\textwidth]{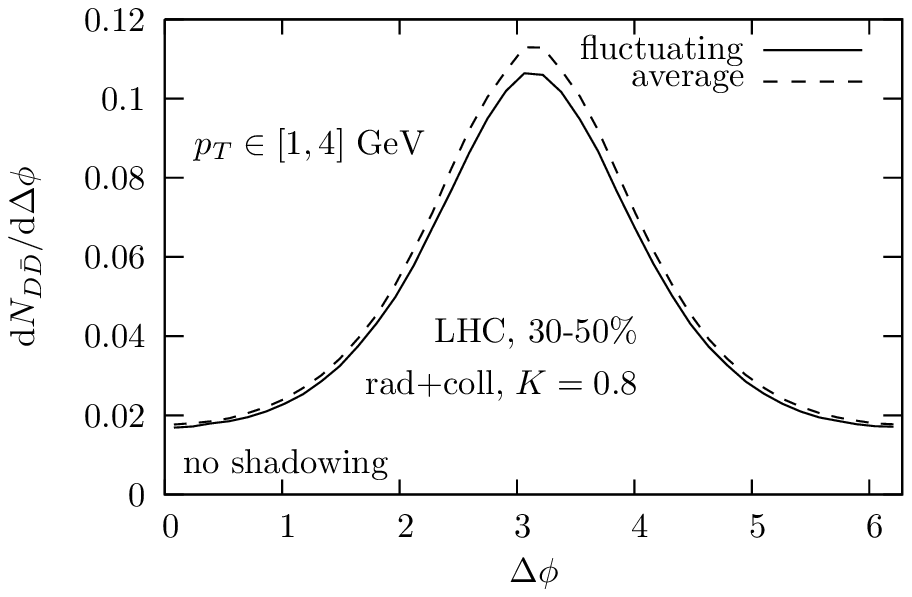}}\hfill
  \subfigure{\label{fig:correl2}\includegraphics[width=0.48\textwidth]{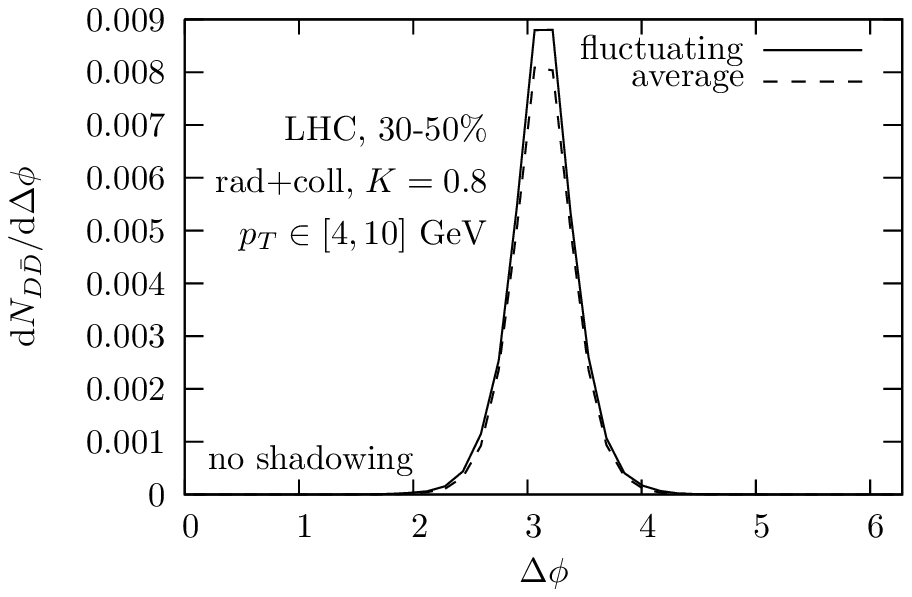}}
 \caption{The azimuthal correlation function of $D\bar{D}$-pairs in two $p_T$-trigger classes compared for fluctuating and averaged initial conditions.}
 \label{fig:correlDDbar}
\end{figure}

Finally in fig.~\ref{fig:correlDDbar}, we show the azimuthal correlations of $D\bar{D}$-pairs, which are initially correlated back to back. These observables do not show any sensitivity to whether the initial conditions are smooth or fluctuating.

\section{Conclusions}

Within our coupled model, MC@sHQ+EPOS, we studied charm quark propagation in non-central heavy-ion collisions at the LHC. We found the expected ordering of all-angle, in- and out-of-plane $R_{\rm AA}$. Analysing the $R_{AA}$, we found that a fluid dynamical expansion from fluctuating 
initial conditions leads to a reduced effective energy loss as compared to the equivalent case with averaged initial conditions.
The quantitative difference between these two scenarios is, however, much smaller than the resolution which can be expected in the experimental data. Our results are qualitatively different from the ones found in \cite{Cao:2014fna}. This shows that different implementation of the energy loss of heavy quarks, of the background fluid dynamical medium and the coupling between these two sectors might put different emphasis on the influence of the initial production points of the heavy quarks or of the fluctuations in the medium during the evolution, as we discussed. It would additionally be interesting to investigate if the different path length dependence of energy loss mechanisms probes the fluctuations in a medium differently. We found that while the elliptic flow at low momenta is reduced in a fluctuating medium it deserves a more detailed further analysis at higher momenta. This is left for when our model is improved to include the consistent initialization of the charm quarks within the EPOS multiple scattering approach and a viscous fluid dynamical evolution.

\medskip

% \section*{Acknowledgements}
MN appreciates fruitful discussions with Steffen Bass and Shanshan Cao. 
We are grateful for support from  the Hessian LOEWE initiative Helmholtz International 
Center for FAIR, ANR research program ``hadrons @ LHC''  (grant ANR-08-BLAN-0093-02),  TOGETHER project R\'egion Pays de la Loire, I3-HP and the U.S. department of Energy under grant DE-FG02-05ER41367.

%% References


% \bibliographystyle{elsarticle-num}
% \bibliography{<your-bib-database>}

\begin{thebibliography}{00}

\bibitem{rhic}
% \bibitem{Stare}
STAR Collab., \textit{Phys. Rev. Lett.} {\bf 98} (2007) 192301; Erratum-ibid. {\bf 106} (2011) 159902;
% \bibitem{Phenixe}
PHENIX Collab.,\textit{ Phys. Rev. C} {\bf 84} (2011) 044905;
%   \bibitem{StarD}
  Dong X [STAR Collaboration],
  %``Highlights from STAR,''
  \textit{Nucl.\ Phys.\ A}904-905 {\bf 2013} (2013) 19c.
%   [arXiv:1210.6677 [nucl-ex]].


\bibitem{Alice} ALICE Collab., \textit{JHEP} {\bf 09} (2012) 112;
%\cite{Dainese:2012ae}
% \bibitem{Dainese:2012ae}
  A.~Dainese [ALICE Collaboration],
  %``Suppression of high-$p_{T}$ heavy-flavour particles in Pb-Pb collisions at the LHC, measured with the ALICE detector,''
  PoS ICHEP {\bf 2012} (2013) 417.
%   [arXiv:1212.0995 [nucl-ex]].
\bibitem{delValle:2012qw}
  Conesa del Valle Z [ALICE Collaboration],
  %``Heavy-flavor suppression and azimuthal anisotropy in Pb-Pb collisions at $\sqrt(s_{NN}) = 2.76$ TeV with the ALICE detector,''
  \textit{Nucl.\ Phys.\ A}904-905 {\bf 2013} (2013) 178c.
%   [arXiv:1212.0385 [nucl-ex]].
 \bibitem{Abelev:2013lca}
   B.~Abelev {\it et al.}  [ALICE Collaboration],
  %``D meson elliptic flow in non-central Pb-Pb collisions at $\sqrt{s_{NN}} = 2.76$TeV,''
  Phys.\ Rev.\ Lett.\  {\bf 111} (2013) 102301.
%   [arXiv:1305.2707 [nucl-ex]].

\bibitem{radiative} Gyulassy M and Wang X N, Nucl. Phys. B {\bf 420} (1994) 583;
% \bibitem{Wang95} 
Wang X N, Gyulassy M, and Pl\"umer M, \textit{Phys. Rev. D} {\bf 51} 
 (1995) 3436;
% \bibitem{Baier95} 
Baier R, Dokshitzer Y L, Peign\'{e} S, and Schiff D, 
 \textit{Phys. Lett. B} {\bf 345} (1995) 277;
% \bibitem{Baier97} 
Baier R, Dokshitzer Y L, M\"uller A H, Peign\'{e} S, and Schiff D, 
 \textit{Nucl. Phys. B} {\bf 483} (1997) 291; \textit{Nucl. Phys. B} {\bf 484} (1997) 265; 
% \bibitem{Zakharov} 
Zakharov B G, \textit{JETP Lett.} {\bf 63} (1996) 952; \textit{JETP Lett.} {\bf 64} (1996) 781; 
 \textit{JETP Lett.} {\bf 65} (1997) 615; \textit{JETP Lett.} {\bf 73} (2001) 49; \textit{JETP Lett.} {\bf 78} (2003) 759; \textit{JETP Lett.} {\bf 80} (2004) 617;
% \bibitem{GLV} 
Gyulassy M, Levai P, and Vitev I, \textit{Phys. Rev. Lett.} {\bf 85} (2000) 
 5535; \textit{Nucl. Phys. B} {\bf 571} (2000) 197; \textit{Nucl. Phys. B} {\bf 594} (2001) 371; 
% \bibitem{Dokshitzer} 
Dokshitzer Y L and Kharzeev D E, \textit{Phys. Lett. B} {\bf 519} 
 (2001) 199; 
% \bibitem{AMY} 
Arnold P B, Moore G D, and Yaffe L G, \textit{JHEP} 0011 (2000) 001; 
 \textit{JHEP} 0305 (2003) 051; 
% \bibitem{ASW} 
Armesto N, Salgado C A, and Wiedemann U A, \textit{Phys. Rev. D} {\bf 69} (2004) 
 114003; \textit{Phys. Rev. C} {\bf 72} (2005) 064910; 
% \bibitem{Zhang04}
Zhang B W, Wang E and Wang X N, 
% Heavy Quark Energy Loss in Nuclear Medium
 \textit{Phys. Rev. Lett.} {\bf 93} (2004) 072301.
% [arXiv :nucl-th/0309040]

\bibitem{elastic} Bjorken J D, Fermilab preprint Pub-82/59-THY (1982);
% \bibitem{Peshier:2006hi}
  Peshier A,
  %``The QCD collisional energy loss revised,''
  \textit{Phys.\ Rev.\ Lett.}  {\bf 97} (2006) 212301;
%\cite{Peigne:2008nd}
% \bibitem{Peigne:2008nd}
  Peigne S and Peshier A,
  %``Collisional energy loss of a fast heavy quark in a quark-gluon plasma,''
  \textit{Phys.\ Rev.\ D }{\bf 77} (2008) 114017.
%   [arXiv:0802.4364 [hep-ph]].

\bibitem{results}
% \bibitem{Moore:2004tg}
  Moore G D and Teaney D,
  %``How much do heavy quarks thermalize in a heavy ion collision?,''
  \textit{Phys.\ Rev.\ C }{\bf 71} (2005) 064904;
%   [hep-ph/0412346].
% \bibitem{Uphoff:2010sh}
  Uphoff J, Fochler O, Xu Z and Greiner C,
  %``Heavy quark production at RHIC and LHC within a partonic transport model,''
  \textit{Phys.\ Rev.\ C} {\bf 82} (2010) 044906;
%   [arXiv:1003.4200 [hep-ph]]
% 
% \bibitem{Alberico:2011zy}
  Alberico W M, Beraudo A, De Pace A, Molinari A, Monteno M, Nardi M and Prino F,
  %``Heavy-flavour spectra in high energy nucleus-nucleus collisions,''
  \textit{Eur.\ Phys.\ J.\ C} {\bf 71} (2011) 1666;
%   [arXiv:1101.6008 [hep-ph]].
% \bibitem{Uphoff:2011ad}
  Uphoff J, Fochler O, Xu Z and Greiner C,
  %``Elliptic Flow and Energy Loss of Heavy Quarks in Ultra-Relativistic heavy Ion Collisions,''
  \textit{Phys.\ Rev.\ C} {\bf 84} (2011) 024908;
%   [arXiv:1104.2295 [hep-ph]].
% \bibitem{Lang:2012cx}
  Lang T, van Hees H, Steinheimer J and Bleicher M,
  %``Heavy quark transport in heavy ion collisions at RHIC and LHC within the UrQMD transport model,''
  arXiv:1211.6912 [hep-ph];
%\cite{Cao:2013ita}
% \bibitem{Cao:2013ita}
    S.~Cao, G.~-Y.~Qin and S.~A.~Bass,
  %``Heavy quark dynamics and hadronization in ultra-relativistic heavy-ion collisions: collisional versus radiative energy loss,''
  Phys.\ Rev.\ C {\bf 88} (2013) 044907.
%   [arXiv:1308.0617 [nucl-th]].

\bibitem{Gossiaux:2008jv}
  Gossiaux P B and Aichelin J,
  %``Towards an understanding of the RHIC single electron data,''
  \textit{Phys.\ Rev.\ C }{\bf 78} (2008) 014904.

\bibitem{Gossiaux:2010yx}
  Gossiaux P B, Aichelin J, Gousset T and Guiho V,
  %``Competition of Heavy Quark Radiative and Collisional Energy Loss in Deconfined Matter,''
  \textit{J.\ Phys.\ G} {\bf 37} (2010) 094019.
%   [arXiv:1001.4166 [hep-ph]].

\bibitem{Gossiaux:2012ya}
  Gossiaux P B, Nahrgang M, Bluhm M, Gousset T and Aichelin J,
  %``Heavy quark quenching from RHIC to LHC and the consequences of gluon damping,''
\textit{Nucl. Phys. A} {\bf 904-905} (2013) 992.
%   arXiv:1211.2281 [hep-ph].

\bibitem{Gossiaux:2011ea}
  Gossiaux P B, Vogel S, van Hees H, Aichelin J, Rapp R, He M and Bluhm M,
  %``The Influence of bulk evolution models on heavy-quark phenomenology,''
  %Submitted to: Phys.rev.C
  [arXiv:1102.1114 [hep-ph]].

%\cite{Nahrgang:2013xaa}
\bibitem{Nahrgang:2013xaa}
    M.~Nahrgang, J.~Aichelin, P.~B.~Gossiaux and K.~Werner,
  %``Influence of hadronic bound states above $T_c$ on heavy-quark observables in Pb+Pb collisions at at the CERN Large Hadron Collider,''
  Phys.\ Rev.\ C {\bf 89} (2014) 014905.

\bibitem{Nahrgang:2013saa}
  M.~Nahrgang, J.~Aichelin, P.~B.~Gossiaux and K.~Werner,
  %``Azimuthal correlations of heavy quarks in Pb+Pb collisions at LHC ($\sqrt{s}=2.76$ TeV),''
  arXiv:1305.3823 [hep-ph].

\bibitem{Nahrgang:2013pka}
  M.~Nahrgang, J.~Aichelin, P.~B.~Gossiaux and K.~Werner,
  %``Heavy-flavor azimuthal correlations of D mesons,''
  arXiv:1310.2218 [hep-ph].

%\cite{Rodriguez:2010di}
\bibitem{Rodriguez:2010di}
  R.~Rodriguez, R.~J.~Fries and E.~Ramirez,
  %``Event-by-Event Jet Quenching,''
  Phys.\ Lett.\ B {\bf 693} (2010) 108
  [arXiv:1005.3567 [nucl-th]].
  %%CITATION = ARXIV:1005.3567;%%
  %17 citations counted in INSPIRE as of 03 May 2014fna

%\cite{Renk:2011qi}
\bibitem{Renk:2011qi}
  T.~Renk, H.~Holopainen, J.~Auvinen and K.~J.~Eskola,
  %``Energy Loss in a Fluctuating Hydrodynamical Background,''
  Phys.\ Rev.\ C {\bf 85} (2012) 044915
  [arXiv:1105.2647 [hep-ph]].
  %%CITATION = ARXIV:1105.2647;%%
  %19 citations counted in INSPIRE as of 03 May 2014

%\cite{Zhang:2012ik}
\bibitem{Zhang:2012ik}
  H.~Zhang, T.~Song and C.~M.~Ko,
  %``Effects of initial state fluctuations on jet energy loss,''
  Phys.\ Rev.\ C {\bf 87} (2013) 5,  054902
  [arXiv:1208.2980 [hep-ph]].
  %%CITATION = ARXIV:1208.2980;%%
  %2 citations counted in INSPIRE as of 03 May 2014

%\cite{Cao:2014fna}
\bibitem{Cao:2014fna}
  S.~Cao, Y.~Huang, G.~-Y.~Qin and S.~A.~Bass,
  %``The Influence of Initial State Fluctuations on Heavy Quark Energy Loss in Relativistic Heavy-ion Collisions,''
  arXiv:1404.3139 [nucl-th].
\bibitem{EPOS}
  Werner K, Karpenko I, Pierog T, Bleicher M and Mikhailov K,
  %``Event-by-Event Simulation of the Three-Dimensional Hydrodynamic Evolution from Flux Tube Initial Conditions in Ultrarelativistic Heavy Ion Collisions,''
  \textit{Phys.\ Rev.\ C} {\bf 82} (2010) 044904;
  Werner K, Karpenko I, Bleicher M, Pierog T and Porteboeuf-Houssais S,
  %``Jets, Bulk Matter, and their Interaction in Heavy Ion Collisions at Several TeV,''
  \textit{Phys.\ Rev.\ C} {\bf 85} (2012) 064907;
  Werner K, Bleicher M, Guiot B, Karpenko I and Pierog T,
  %``Evidence for flow in pPb collisions at 5 TeV from v2 mass splitting,''
  arXiv:1307.4379 [nucl-th].
%   [arXiv:1203.5704 [nucl-th]].
  %%CITATION = ARXIV:1203.5704;%%

\bibitem{fonll}
  Cacciari M, Greco M and Nason P,
  %``The p(T) spectrum in heavy-flavour hadroproduction,''
  \textit{JHEP} {\bf 9805} (1998) 007;
%   [arXiv:hep-ph/9803400];
  %%CITATION = HEP-PH 9803400;%%
% \bibitem{FONLL2}
  Cacciari M, Frixione S and Nason P,
  %``The p(T) spectrum in heavy-flavor photoproduction,''
  \textit{JHEP} {\bf 0103} (2001) 006;
%   [arXiv:hep-ph/0102134].
% \bibitem{FONLL3}
  Cacciari M, Frixione S, Houdeau N, Mangano M L, Nason P and Ridolfi G,
  %``Theoretical predictions for charm and bottom production at the LHC,''
  \textit{JHEP} {\bf 1210} (2012) 137;
%   [arXiv:1205.6344 [hep-ph]].
% \bibitem{fonllform}
www.lpthe.jussieu.fr/~cacciari/fonll/fonllform.html.
%\cite{Gossiaux:2012cv}
\bibitem{Gossiaux:2012cv}
  P.~B.~Gossiaux,
  %``Recent results on heavy quark quenching in ultrarelativistic heavy ion collisions: the impact of coherent gluon radiation,''
%   arXiv:1209.0844 [hep-ph].
Nucl.\ Phys.\ A 910-911 (2013), 301-305.
  %%CITATION = ARXIV:1209.0844;%%
  %6 citations counted in INSPIRE as of 03 May 2014


 \end{thebibliography}



\end{document}